# The role of the cooperative Jahn-Teller effect in the charge ordered La$_{1-x}$Ca$_x$MnO$_3$ (0.5≤$x$≤0.87) manganites

R. K. Zheng[1], G. Li[1], A. N. Tang[1], Y. Yang[1], W. Wang[1], X. G. Li[1,a)],

Z. D. Wang[1,2], H. C. Ku[3]

1. Structure Research Laboratory, Department of Materials Science and Engineering, University of Science and Technology of China, Anhui, Hefei 230026, China
2. Department of Physics, The University of Hong Kong, Hong Kong, China
3. Department of Physics, The National Tsinghua University, Hsinchu 300, Taiwan, China

Based on the magnetoresistance, magnetization, ultrasound, and crystallographic data, we studied the role of the cooperative Jahn-Teller effect in the charge ordered (CO) state for La$_{1-x}$Ca$_x$MnO$_3$. We found that, with increasing the fraction of $Q_3$ mode of Jahn-Teller distortion and decreasing that of $Q_2$ mode in the CO state, the magnetic structure evolves from CE-type to C-type and the orbital ordering changes from $3d_{x^2-r^2}/3d_{y^2-r^2}$-type to $3d_{z^2-r^2}$-type, with the strength of ferromagnetism and the phase separation tendency being suppressed. At the same time, the stability of the CO state and the cooperative Jahn-Teller lattice distortion increase. These effects imply that the cooperative Jahn-Teller effect with different vibration modes is the key ingredient in understanding the essential physics of the CO state.

---

a) Corresponding author, Electronic mail: Lixg@ustc.edu.cn

The charge, spin, and orbital orderings in manganites have recently attracted much attention because of their intriguing electronic, magnetic, and structural properties [1,2]. As a typical example of charge ordering cases, the $La_{1-x}Ca_xMnO_3$ system with the doping range of $0.5 \leq x \leq 0.875$, especially for a commensurate fraction of doping level such as $x=1/2, 2/3, 3/4$, undergoes evident charge, spin, and orbital ordering phase transitions upon cooling below the charge ordering transition temperature $T_{CO}$ [3,4]. In the charge ordered (CO) state, the ordering of $Mn^{3+}$ and $Mn^{4+}$ species within the $MnO_2$ plane leads to exotic static stripe phase with insulator antiferromagnetic ground state [3,5]. With increasing $x$ from 0.5 to 0.75 the magnetic structure evolves gradually from CE-type to C-type, and is almost C-type for $x>0.75$ [6], with the strength of ferromagnetism and phase separation tendency being suppressed [7,8]. At the same time, the magnitude of the cooperative Jahn-Teller distortion and the robustness of the CO state increase [9]. Moreover, the crystal structure in the CO state changes from tetragonally compressed orthorhombic ($b/\sqrt{2} < a \approx c$ for $x<0.75$) to tetragonally elongated orthorhombic ($c \approx a < b/\sqrt{2}$ for $x>0.75$) near $x=0.75$ [10]. For $x>0.75$, both the Jahn-Teller distortion and the stability of CO state decrease with increasing doping level $x$. Although theoretical and experimental studies have indicated that the magnetic, transport, and crystallographic properties, the phase separation, and the electron-lattice interaction in manganites are closely related to the Jahn-Teller effect [2,11-14], an *overall* profound understanding on the key role of the Jahn-Teller effect in the CO state and detailed explanations for the observed experimental results are still incomplete.

In this Letter, we *experimentally* elucidate the relationships of the cooperative Jahn-Teller effect having different vibration modes with a variety of factors in $La_{1-x}Ca_xMnO_3$, such as the evolutions of magnetic and crystal structures, the orbital ordering, the stability of the CO state, the strength of ferromagnetism, the phase separation tendency, the magnitude of the lattice distortion. The analyses of the experimental data suggest that the cooperative Jahn-Teller effect with different vibration mode plays a key role in understanding the rich physics of the charge ordered state.

The $La_{1-x}Ca_xMnO_3$ samples were synthesized via a coprecipitation method.

Resistivity $\rho$(T) was measured using a standard four-probe technique in magnetic fields up to 14T. Magnetizations were measured at H=0.1T and 5T using a Quantum Design SQUID magnetometer. The longitudinal ultrasound was measured using the Matec-7700 series ultrasonic equipment using a pulse-echo-overlap technique.

Based on our previous systematic resistivity data measured under magnetic fields up to 14T for $La_{1-x}Ca_xMnO_3$ (0.5≤x≤0.9) [9] series, we here analyze the magnetoresistance [MR=$(\rho_0 - \rho_H)/\rho_H$] effect under the field of 14T at 75K for different doping levels. The $x$-dependence of the MR effect is plotted in Fig. 1(a). The change of $T_{CO}$ induced by the field(14T), i.e. $\Delta T_{CO} = T_{CO}(0T) - T_{CO}(14T)$, is also plotted against $x$ in the inset of Fig. 1(a). It is seen that the MR effect decreases rapidly from the order of $10^5$ to nearly zero with increasing $x$ from 0.5 to 0.75, while it increases slightly with further increasing $x$ from 0.75 to 0.87. The $x$-dependence of $\Delta T_{CO}$ shows a very similar behavior as the MR does. These magnetic field effects on MR and $\Delta T_{CO}$ demonstrate that the CO state becomes increasingly robust as $x$ increases from 0.5 to 0.75, and appears to be the most robust at (or near) $x$=0.75. One may naturally think that these magnetic field effects on MR and $\Delta T_{CO}$ may relate to the intrinsic strength of the ferromagnetism of the system. To clarify this point, we measured the temperature dependence of the magnetization at H=0.1T and 5T for $La_{1-x}Ca_xMnO_3$ series. As shown in the inset of Fig. 1(b), the magnetization shows a peak at $T_{CO}$, and decreases prominently with the development of charge and orbital ordering state. It is interesting to find that the magnetization at $T_{CO}$ (i.e. $M_{T_{CO}}$) is the largest for $x$=0.5, and decreases rapidly with increasing $x$ from 0.5 to 0.75, reaching the minimum at $x$≈0.75. When $x$ increases from 0.75 to 0.9 the $M_{T_{CO}}$ increases slightly. By comparing the $x$-dependences of the MR and the $\Delta T_{CO}$ with that of the $M_{T_{CO}}$, it is straightforward to conclude that the decrease of the MR and the $\Delta T_{CO}$ with increasing $x$ is due to the decrease of intrinsic strength of the ferromagnetism of the system. However, what drives the system into less ferromagnetic state as $x$ increases from 0.5 to 0.75 ?

It has been shown that the ratio of Jahn-Teller vibration mode $Q_3/Q_2$ in a given static distortion is [15-17]

$$\tan\Phi = \frac{(2/\sqrt{6})(2m-l-s)}{\pm(2/\sqrt{2})(l-s)}, \qquad (1)$$

where $s$ and $l$ are the short and long Mn-O bond lengths pointing along the [100] and [010] axes alternatively, $m$ is the bond length along the [001] axis, $s \leq m \leq l$, and $\Phi$ is the angle between the state vectors and the $Q_2$ axis, which measures the relative fractions of $Q_3$ and $Q_2$ modes. In the high-anisotropy limit, $m=s$, only the $Q_3$ mode is present, corresponding to $\Phi=30°$; while for the low-anisotropy case $\Phi=0°$, there is only $Q_2$ mode[17]. In fact, in real Jahn-Teller distorted materials, both the $Q_2$ and $Q_3$ modes contribute to the lattice distortion, nevertheless, the ratio $Q_3/Q_2$ depends on the value of $l$, $m$, and $s$. Using the Mn-O bond length obtained from low temperature high resolution neutron diffraction measurements for $La_{0.5}Ca_{0.5}MnO_3$ [5], $La_{0.33}Ca_{0.67}MnO_3$ [18], $La_{1-x}Ca_xMnO_3$ ($x=0.8, 0.85$) [19], and the Mn-O bond length at 70 K obtained from Rietveld analysis of our high resolution powder x-ray diffraction data for $x=0.75$ ($R_P=8.14\%$, $R_{WP}=11.4\%$, $\chi^2=1.74$), we calculated the $\Phi$ using Eq.(1). The $x$-dependence of the $\Phi$ is plotted in Fig. 2(a). It is seen that, for $x=0.5$, $\Phi=13.5°$, indicating a modest anisotropy. Interestingly, when $x$ increases from 0.5 to 0.75, the $\Phi$ increases almost linearly and evolves toward high-anisotropy limit ($\Phi=24.53°$ for $x=0.75$). The increase of $\Phi$ suggests a growth of $Q_3$ mode and a decrease of $Q_2$ mode. With further increase of $x$ from 0.75, the $\Phi$ decreases slightly, but it remains at a high value, indicating that the $Q_3$ is the predominant mode for $x>0.75$. If we assume that the fraction of $Q_3$ mode in the high-anisotropy limit (i.e. $\Phi=30°$, corresponding to pure $Q_3$ mode) is 1, then the fraction of $Q_3$ mode in the CO state at a fixed doping level $x$ can be calculated as

$$Q_3\% = \frac{\tan\Phi°}{\tan 30°}, \qquad (2)$$

Following this equation, we obtained the $x$-dependence of the fraction of $Q_3$, as shown in the inset of Fig. 2(a). The $Q_3\% \sim x$ relationship clearly demonstrates that the fraction of $Q_3$ mode increases almost linearly with increasing doping level $x$. As we recently reported that the relative change of the ultrasound (i.e. the $\Delta V/V$) reflecting the magnitude of the cooperative Jahn-Teller distortion in the CO state also increases almost linearly with increasing doping level from $x=0.5$ to $x=0.75$ [9]. The $\Delta V/V$ versus $x$ curve is shown in Fig. 2(b). It is, therefore, very possible that the increase of the cooperative Jahn-Teller distortion (or equivalently the $\Delta V/V$) with increasing $x$ can be intrinsically related to the increase of the fraction of $Q_3$ mode. To further look into this issue, we plot the $\Delta V/V$ as a function of $Q_3\%$, as shown in the inset of Fig. 2(b). The $\Delta V/V$ increases almost linearly with the increase of the fraction of $Q_3$, which strongly suggest that the increase of the $\Delta V/V$ is microscopically due to the growth of $Q_3$ mode Jahn-Teller distortion and decrease of $Q_2$ mode, and moreover, demonstrate that the cooperative Jahn-Teller distortion contributed from $Q_3$ mode is much more prominent than that contributed from $Q_2$ mode.

It is worth noting that the neutron diffraction measurements on $La_{1-x}Ca_xMnO_3$ ($0.5 \leq x \leq 1$) have shown that the magnetic structure in the ground state for $x=0.5$ is CE-type, while is C-type for $x>0.75$ [6]. For $0.5<x<0.75$, the magnetic structure is a mixture of the two types, and the C-type magnetic structure grows at the cost of the CE-type's decline with increasing $x$ from 0.5 to 0.75 [6]. As pointed out in Refs.[2,11-14], the CE-type magnetic structure originates from $3d_{x^2-r^2}/3d_{y^2-r^2}$ orbital ordering consistent with $Q_2$ mode cooperative Jahn-Teller distortion and the C-type magnetic structure originates from $3d_{z^2-r^2}$ orbital ordering consistent with $Q_3$ mode cooperative Jahn-Teller distortion. Based on this scenario, one can conclude that the growth of C-type magnetic structure with increasing $x$ is the result of the increase of the fraction of $Q_3$ mode cooperative Jahn-Teller distortion, corresponding to the evolution of the orbital ordering from $3d_{x^2-r^2}/3d_{y^2-r^2}$ to $3d_{z^2-r^2}$. From this picture, the phase separation tendency and the magnetic field effects on the stability of the CO state can readily be understood as follows: with increasing $x$ from 0.5 to 0.75 the strength of the electron-lattice interaction with the cooperative Jahn-Teller distortion and the fraction of C-type magnetic structure increase because of the increase of the fraction of the $Q_3$ mode Jahn-Teller distortion. Hence, the strength of the ferromagnetism of the system was suppressed, and thus the phase separation tendency decreases. On the other hand, it is natural to expect that the magnetic field effects on the stability of CO state will become less effective because the ferromagnetism of the system decreases in the presence of more and more C-type magnetic structure. In fact, the cooperative Jahn-Teller effect with different vibration mode on the stability of CO state (or MR) can be more intuitively understood if we plot the $\frac{1}{\Delta V/V}$ as a function of doping level $x$ in Fig. 1(a). The Jahn-Teller effect with different vibration mode effects on the magnetism is also observed for low-doped $R_{2/3}D_{1/3}MnO_3$ (R=La, Tb, La-Pr, Pr, La-Y, D=Ca, Sr) manganites where a low magnetic moment is always associated with the predominance of $Q_3$ mode Jahn-Teller distortion and a high magnetic moment with $Q_2$ mode [17].

In summary, we have studied the role of cooperative Jahn-Teller effect in the CO state for $La_{1-x}Ca_xMnO_3$. The evolutions of magnetic structure from CE-type to C-type, the orbital ordering from $3d_{x^2-r^2}/3d_{y^2-r^2}$ to $3d_{z^2-r^2}$, and crystal structure from tetragonally compressed to tetragonally elongated orthorhombic, the suppression of ferromagnetism and phase separation tendency, the increase of the robustness of the CO state, the increase of the magnitude of the cooperative Jahn-Teller lattice distortion can all be attributed to the increase of the relative fraction of $Q_3$ mode Jahn-Teller distortion with respect to $Q_2$ mode. These results suggest that the cooperative Jahn-Teller effect with different vibration modes plays an essential role of physics in the CO state for the $La_{1-x}Ca_xMnO_3$ system.

This work was supported by the Chinese National Nature Science Fund, and the Ministry of Science and Technology of China.

**Figure Captions**

Fig. 1 (a) The magnetoresistance (MR) effect at T=75 K and H=14T as a function of doping level $x$ for $La_{1-x}Ca_xMnO_3$. The solid line is a guide to the eyes. The open circles are the $\frac{1}{\Delta V/V}$ versus $x$ data. Inset in (a) is the $x$-dependence of the magnetic field (14T) induced change of the $\Delta T_{CO}$. (b) The magnetization at $T_{CO}$ ($M_{T_{CO}}$) measured at H=0.1T and 5T as a function of doping level $x$, the solid line is a guide to the eyes for H=0.1T, the open circles are the $M_{T_{CO}}$ versus $x$ data measured at H=5T. Inset in (b) is the temperature dependence of the zero-field cooled magnetization for $La_{1-x}Ca_xMnO_3$.

Fig. 2 (a) Variation of the Jahn-Teller vibration anisotropy $\Phi$ as a function of doping level $x$. The inset in (a) is the $x$-dependence of the fraction of the $Q_3$ mode. (b) Variation of the relative change of the ultrasound ($\Delta V/V$) as a function of doping level $x$. The solid line is a guide to the eyes. The inset (upper) in (b) is the $\Delta V/V$ versus $Q_3\%$ curve. The inset (lower) in (b) is a typical temperature dependence of the ultrasound for $La_{1-x}Ca_xMnO_3$.

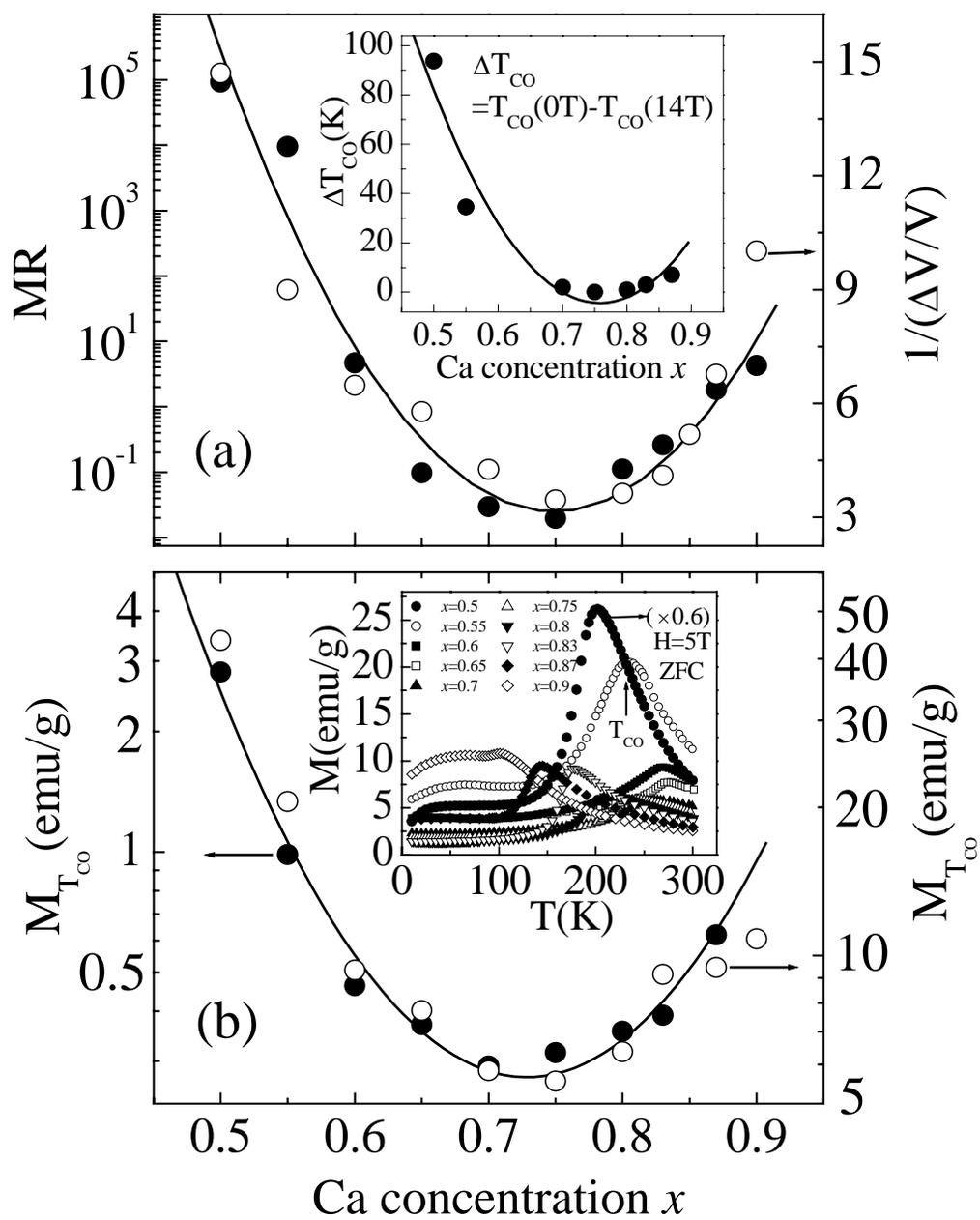

**Fig. 1 By R. K. Zheng et al.**

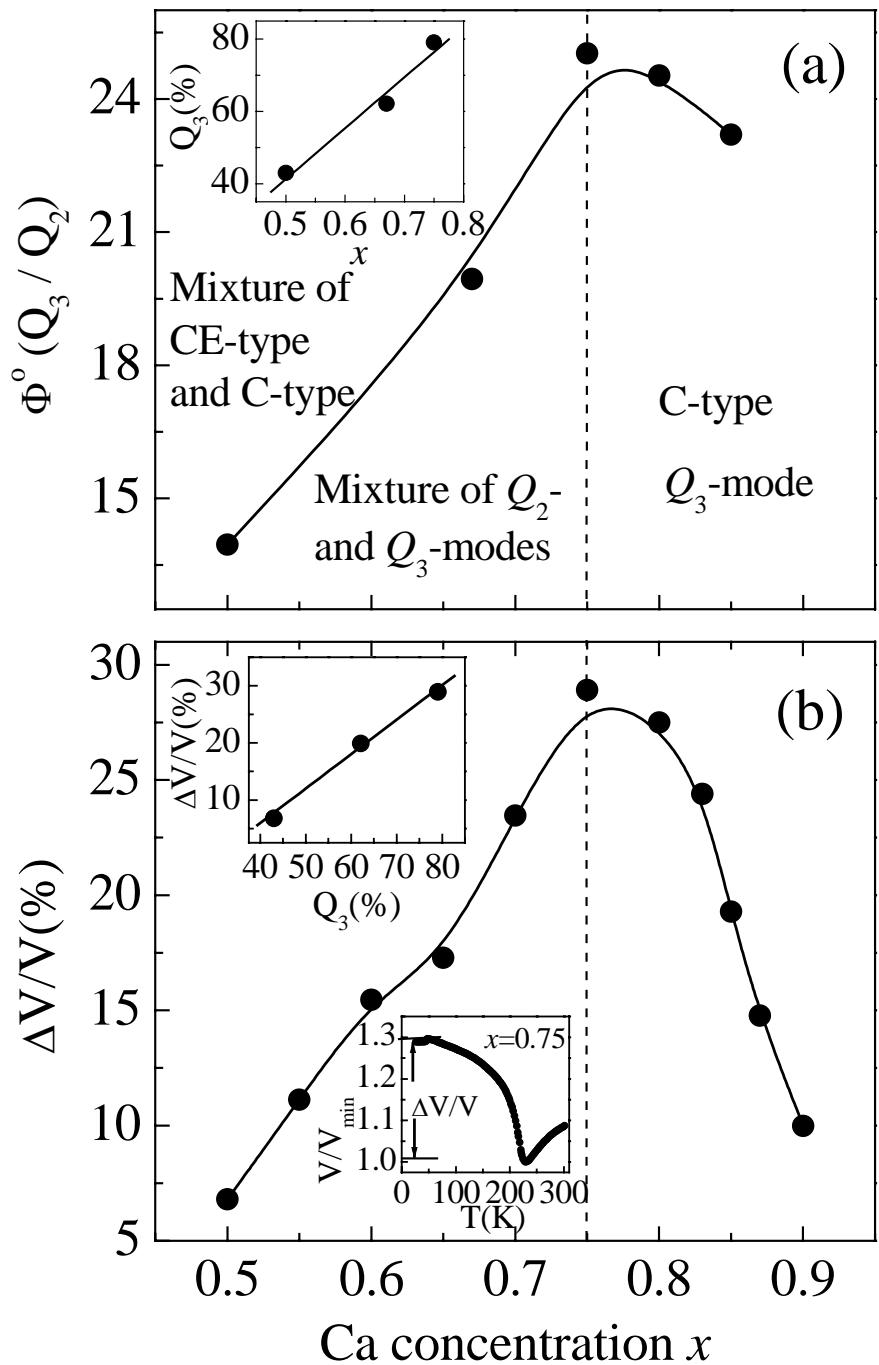

Fig. 2 By R. K. Zheng et al.

10